\documentclass{PoS}

\RequirePackage{graphicx}
\RequirePackage{amsmath}
\usepackage{amssymb}
\usepackage{cite}
\usepackage{bm}
\usepackage{caption}
\usepackage{slashed}
\usepackage{simplewick}
\usepackage{color}
\usepackage{amsthm}

\title{Examining the Low Energy Dynamics \\ of Walking Gauge Theory}

\ShortTitle{Walking EFT}

\author{ Andrew D. Gasbarro \\
	Yale University, Sloane Physics Laboratory, 217 Prospect Street, New Haven, CT 06511 \\
	E-mail:  \email{Andrew.Gasbarro@yale.edu}
}

\author{ George T. Fleming \\
	Yale University, Sloane Physics Laboratory, 217 Prospect Street, New Haven, CT 06511 \\
	E-mail:  \email{George.Fleming@yale.edu}
}

\author{for the Lattice Strong Dynamics (LSD) Collaboration} 

\abstract{We report on an investigation into the low energy dynamics of walking gauge theory.  Taking $SU(3)$ Yang Mills with eight flavors of fundamental fermions as an example, we discuss the light flavor singlet scalar appearing in the spectrum and its implications for low energy physics.  We compute the maximal isospin $\pi\pi$ scattering length at the lightest quark masses yet investigated for the eight flavor theory.  The validity of chiral perturbation theory is assessed, and we discuss motivations for a more extensive effective field theory analysis to be carried out in future work.}

\FullConference{34th annual International Symposium on Lattice Field Theory\\
		24-30 July 2016\\
		University of Southampton, UK}

\begin{document}

\maketitle

\section{Introduction \label{sec:one}}

Yang-Mills gauge theories exhibit a variety of surprising behaviors in the infrared as the number of colors, $N_c$, and number of flavors, $N_F$, are varied.  While Seiberg duality allows one to make analytical progress on understanding the IR phases of supersymmetric gauge theory, the nonsupersymmetric case has primarily been examined using lattice techniques.  In particular, for fermions transforming in the fundamental representation of the gauge group, many studies have been carried out in an attempt to establish the lower boundary of the conformal window, $N_F^*$, at which the theory transitions from conformal to confining behavior in the IR \cite{conformalwindow,conformalwindow2} (see Ref.~\cite{Pica:2017gcb} for a recent review).  Near the conformal window, the gauge coupling is postulated to evolve slowly with scale.  This ``walking'' behavior has been suggested as a solution to some of the maladies that arise when QCD-like strong dynamics are applied to dynamical electroweak symmetry breaking (DEWSB) in technicolor models.  The novel features of walking dynamics include a reduced electroweak S-parameter \cite{Sparameter}, an enhanced chiral condensate allowing for suppression of flavor changing neutral currents in extended technicolor \cite{conformalwindow2}, and a light flavor singlet scalar \cite{8flightsigma1,8flightsigma2,8flightsigma3}.

While DEWSB through strong dynamics is a good motivation for studying walking and conformal Yang Mills, we stress that studying these gauge theory dynamics for their own sake can be of interest to a great variety of areas in high energy theory.  For example, in the excitement of the 750 GeV diphoton ``bump'' at the LHC in 2016 \cite{750GeV1,750GeV2}, some model builders suggested strong dynamics as a mechanism to explain the large production rate of the would-be BSM particle \cite{750model1}.  These models often appealed to intuition from QCD and leading order chiral perturbation theory ($\chi$PT).  However, walking dynamics were also of interest to model builders and could have been utilized more widely if their low energy properties were better established---for example, if the low energy effective field theory (EFT) of a walking gauge theory was known.  


In this work, we report on an investigation into the low energy behavior of $SU(3)$ gauge theory with eight flavors of fermions in the fundamental representation.  This model is believed by many to be on the confining side of the conformal window but very near the edge, though this is still a matter of some debate.  The nonperturbative beta function   appears to be small and nonzero consistent with the walking hypothesis \cite{8frunning,Fodor:2015baa}. A study of the eight flavor theory with domain wall fermions gave evidence of condensate enhancement and a reduction of the S-parameter \cite{8fDW}.  In this work we will proceed under the assumption that the theory is chirally broken.  We extend the investigation of the eight flavor theory by computing new observables which probe the low energy effective theory.  We stress that our aim in continuing to investigate this model is not to exalt the eight flavor theory as a privileged candidate for Higgs compositeness or any other BSM scenario, but rather to establish general features of walking gauge theories by studying the eight flavor theory as an exemplar. We utilize staggered fermions as our lattice discretization for their efficiency when generating large lattices at light quark masses.  Since our model contains a multiple of four flavors, we do not have to consider rooting.  The lattice action includes two degenerate flavors of staggered fermions which would become eight degenerate flavors of Dirac fermions in the continuum limit.  

In our discussion of effective field theory considerations in this work, let us clarify that we will only be attempting to address the question of the correct EFT for the mass degenerate walking theory.  When discussing the low energy physics of walking gauge theory, it is sometimes tempting to speculate about the mass-split system in which explicit mass terms are introduced to lift extra Goldstone boson states not appearing in the Standard Model electroweak sector (for a numerical investigation, see e.g.~\cite{addition1,addition2}).  However, these are ultimately two separate effective field theory analyses: one of the mass degenerate system in which the EFT is constructed around a $SU(N_F)\times SU(N_F) \rightarrow SU(N_F)$ symmetry breaking pattern, and the other of the mass split system in which the effective theory is the Standard Model plus higher dimensional operators.  We take the approach that more progress should be made on the mass degenerate EFT before one complicates the model by introducing symmetry breaking terms.  Indeed, it remains an open question how the spectrum of walking models will be affected by flavor breaking mass terms, and this will likely require further numerical study.  Therefore, in this work we only consider the EFT built around the unbroken global symmetries of the walking model.


In Section~\ref{sec:two} we report on the recent investigation of the spectrum of the eight flavor theory at the lightest quark masses yet investigated.  Section~\ref{sec:three} details a computation of pion scattering length in the maximal isospin channel using Luscher's method.  In Section~\ref{sec:four}, we discuss the implications of these results for the viability of chiral perturbation theory and other EFTs.  In Section~\ref{sec:five} we conclude and discuss ongoing and future work.

\section{The Eight Flavor Spectrum \label{sec:two}}

\begin{figure}[t!]
	\begin{center}
		\includegraphics[width=0.6\textwidth]{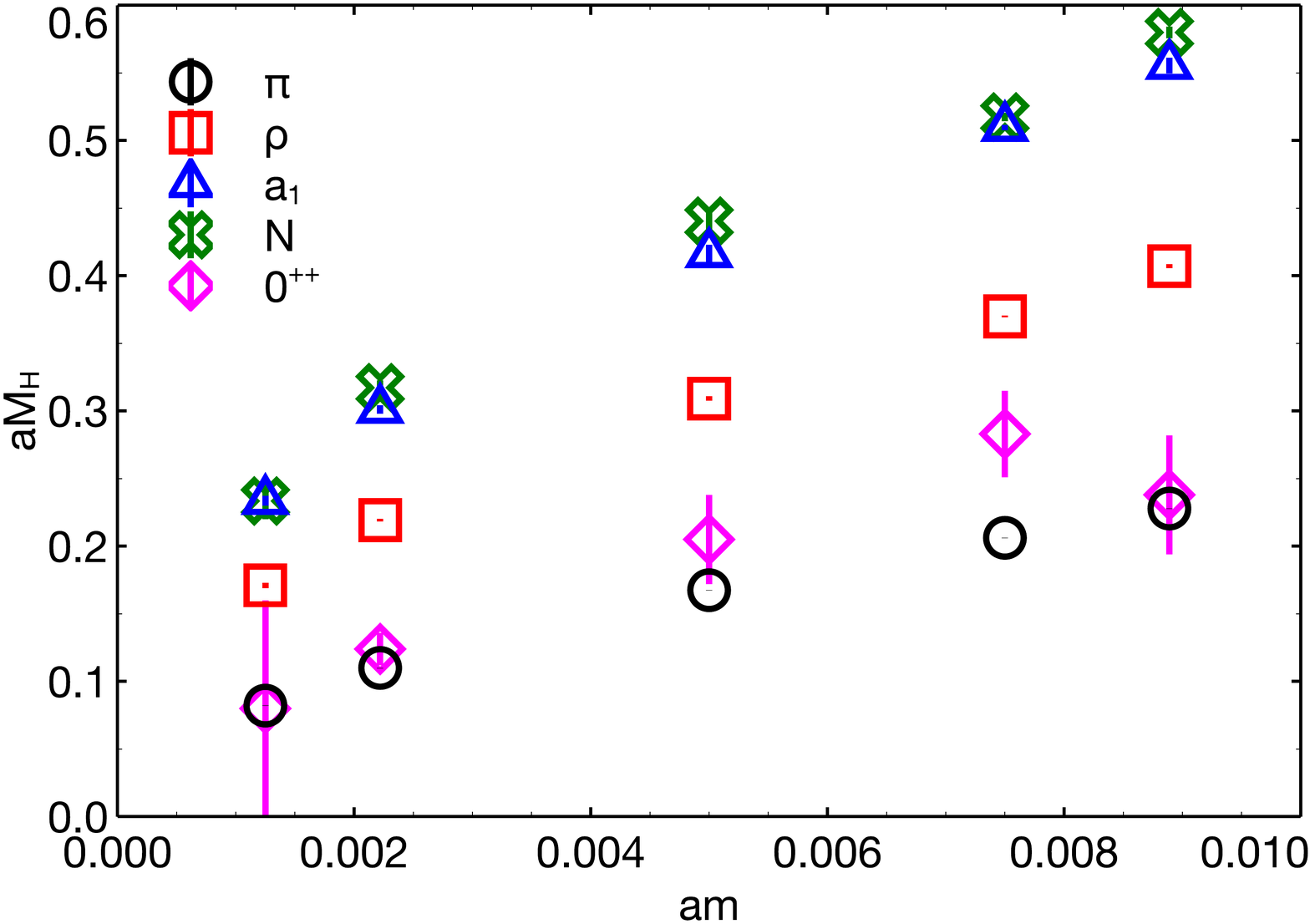}
	\end{center}
	\caption{Spectrum of the eight flavor theory.  The $0^{++}$ error bars include systematic and statistical uncertainties while error bars on the other hadron masses include only statistical errors.}
	\label{fig:spectrum1}
\end{figure}

Here we summarize recent progress made on computations of the hadron spectrum and meson decay constants in the eight flavor theory, with a particular emphasis on the light flavor singlet scalar mass.  It is expected that theories exhibiting softly broken conformal symmetry should possess an additional light state with the quantum numbers of the vacuum, the technidilaton \cite{Yamawaki:1985zg,Bando:1986bg,TAdilaton}.  An initial calculation of the flavor singlet scalar in the eight flavor theory carried out by the latKMI collaboration revealed that the scalar mass is degenerate with the pion and well separated from the $\rho$ mass down to bare quark masses of $a m_q = 0.015$ \cite{8flightsigma1}, with follow up work reaching bare quark masses of $a m_q = 0.012$ \cite{8flightsigma3}.  However, it was unclear how the spectrum would behave as one continued to approach the chiral limit.  The lack of a satisfactory EFT description of the low energy physics in these mass ranges makes it difficult to make definitive statements about chiral extrapolations.  Seeking to gain more insight into the chiral behavior of the eight flavor theory, we have generated ensembles with significantly lighter physical quark masses.  In a recent publication, we have reported measurements of the low lying spectrum including the $0^{++}$ mass\cite{8flightsigma2}.  Here, we record some updated results and point out features of the spectrum that encourage a deeper study of the low energy effective theory of this model.   

\begin{figure}[t!]
	\begin{center}
		\includegraphics[width=0.6\textwidth]{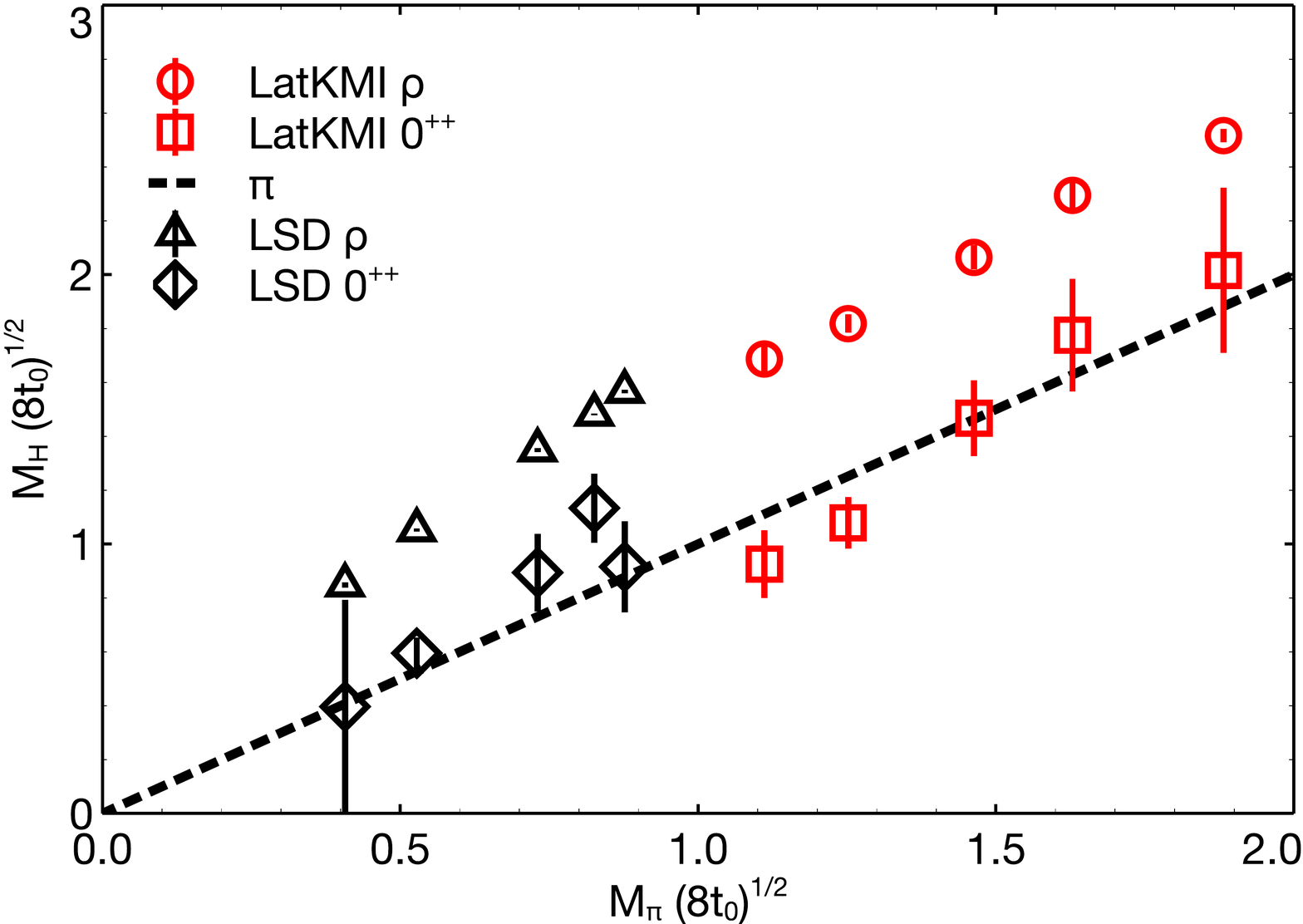}
	\end{center}
	\caption{Combined data from LSD collaboration\cite{8flightsigma2} and latKMI collaboration  \cite{8flightsigma1,8flightsigma3} comparing $M_\rho$ and $M_{0^{++}}$ against $M_\pi$.  Dimensionful quantites are given in units of the Wilson flow scale.}
	\label{fig:spectrum2}
\end{figure}

Fig.~\ref{fig:spectrum1} shows our analysis of the $N_F=8$ spectrum.  Bare quark masses and lattice volumes range from a heaviest quark mass of $a m_q = 0.00889$ on a $24^3\times 48$ lattice to a lightest quark mass of $a m_q = 0.00125$ on a $64^3 \times 128$ lattice.  In Fig.~\ref{fig:spectrum2} we show the combined data from our study and previous studies for the $0^{++}$ and $\rho$ masses \cite{8flightsigma1,8flightsigma2,8flightsigma3}, which demonstrates our approach to the chiral limit. The spectrum shows a clear deviation from QCD in the behavior of $M_{0^{++}}$, which is degenerate with $M_\pi$ within errors and is well separated from $M_\rho$.  If we assume that the lightness of the $0^{++}$ is arising from the broken conformal symmetry, then we should expect the singlet mass to break off from the pion mass at some yet smaller value of $a m_q$.  That is, since the conformal symmetry is explicitly broken by the beta function we would expect the $0^{++}$ to become a pseudotechnidilaton with nonzero mass in the chiral limit rather than an exactly massless technidilaton.  The pions, of course, are exact Nambu-Goldstone bosons in the chiral limit.  It is remains unclear how close to the chiral limit one must go before the splitting can be seen.  We will return to this point in Section~\ref{sec:four}.  

\begin{figure}[t!]
	\begin{center}
		\includegraphics[width=0.7\textwidth]{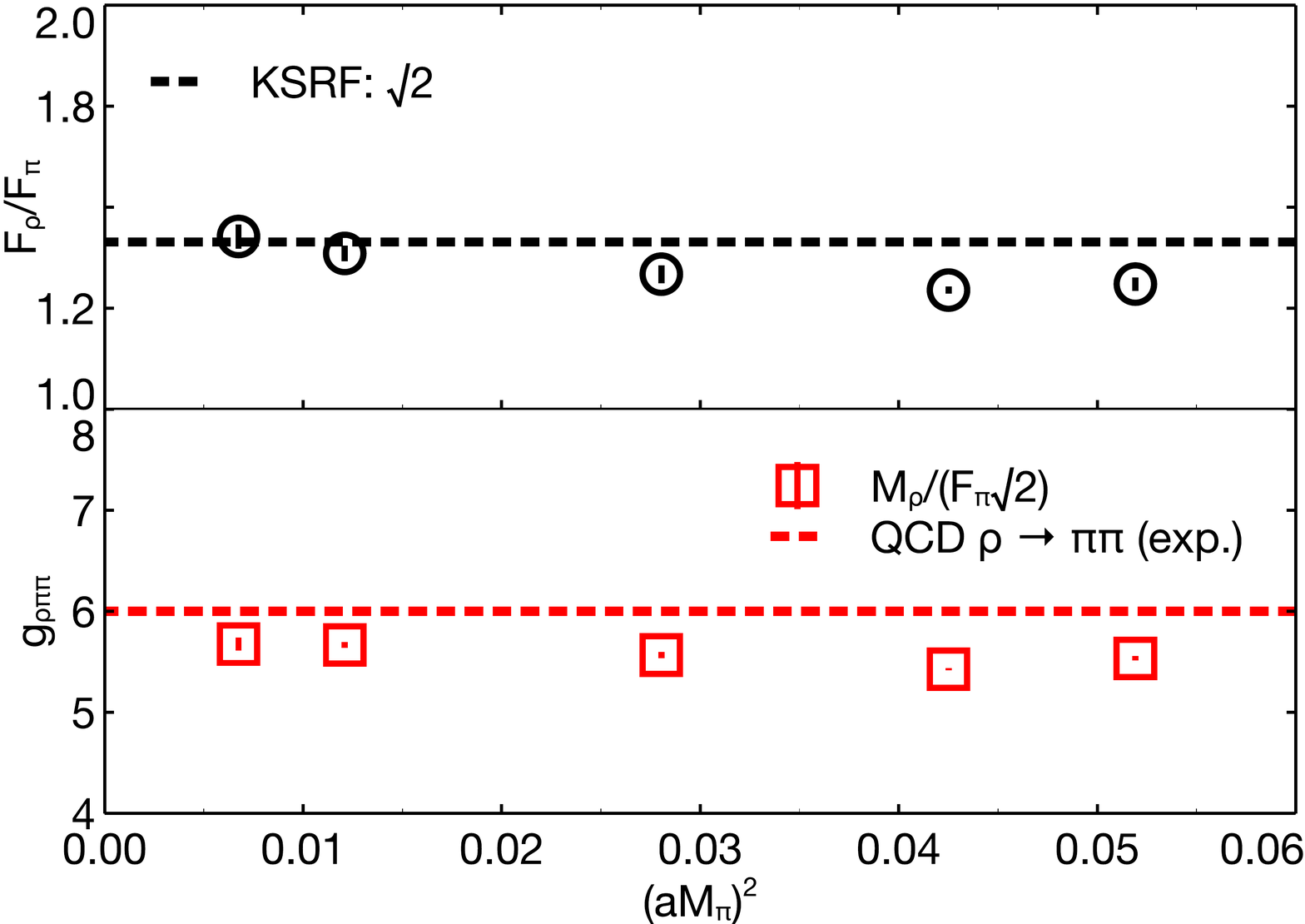}
	\end{center}
	\caption{Tests of the first (upper) and second (lower) KSRF relations in the eight flavor theory.  The lower panel also demonstrates that $M_\rho / F_\pi \approx 8$ and is approximately constant over the mass range studied.}
	\label{fig:KSRF}
\end{figure}

Despite the eight flavor theory's significant deviation from QCD in the singlet scalar mass, other quantities remain surprisingly unchanged.  We test the KSRF relations which hold in QCD within $10\%$ and follow from the assumption of vector meson dominance (VMD).  
\begin{equation}
F_\rho = \sqrt{2} F_\pi  \qquad \qquad g_{\rho\pi\pi} = \frac{M_\rho}{\sqrt{2} F_\pi}
\end{equation}
In Fig.~\ref{fig:KSRF}, the upper panel demonstrates that the ratio $F_\rho / F_\pi$ agrees to within $8\%$ with the KSRF prediction of $\sqrt{2}$.  This gives us some confidence that the second KSRF relation will also hold.  In the lower panel, we plot the predicted value of $g_{\rho\pi\pi}$ from the second KSRF relation.  We find that this agrees with the QCD value $g_{\rho\pi\pi} \approx 6$ (or equivalently, $M_\rho / F_\pi \approx 8$) to within $10\%$.  These results suggest that VMD may be a generic feature of all confining Yang Mills theories both near to and far from the conformal window.  Then, in a composite Higgs scenario in which the scale is set at $F_\pi \approx 250\text{GeV}$, one would expect to find new vector resonances near 2 TeV.  In 2015, some excitement occurred when an excess of this kind was observed at ATLAS and CMS \cite{2TeV1,2TeV2}.  Though no new state was discovered, this does not preclude future discoveries of states of this kind.  Indeed, if one estimates the width $\Gamma_\rho$ using the KSRF relations, 
\begin{equation}
\Gamma_\rho \approx \frac{g_{\rho\pi\pi}^2 M_\rho}{48 \pi} \approx \frac{M_\rho^3}{96 \pi F_\pi^2}
\end{equation}
ones find $\Gamma_\rho \approx 450 \text{ GeV}$ for $M_\rho \approx 2 \text{ TeV}$, and such a wide resonance would be difficult to discover at the LHC.  We are currently carrying out an investigation of the vector form factor of the pion in the eight flavor theory in order to directly study $g_{\rho\pi\pi}$ and the VMD scenario.  In future work, we also plan to investigate the $I=1$ $\pi\pi$ scattering channel in which the $\rho$ is resonant.  For further information about the study of the eight flavor spectrum, see \cite{8flightsigma2}.

While the similarities between the walking theory and QCD are intriguing, the biggest mystery remains how the $0^{++}$ state will behave closer to the chiral limit.  In QCD, one typically considers the $\sigma$ mass to set the radius of convergence of chiral perturbation theory.  The lightness of the flavor singlet scalar suggests some new EFT should apply which encapsulates the dynamics of the Goldstone bosons and scalar (or possibly, scalars) up to the next heaviest state, the $\rho$ mass.  At sufficiently light quark masses, $M_{0^{++}}/M_\pi$ should increase above one and in this regime a low energy description of only pions by a chiral Lagrangian may well hold up to energies $q^2 \approx M_{0^{++}}^2$.  But, pressing down to ever lighter quark masses comes with great computational expense.  Furthermore, if we were in a regime where  $M_{0^{++}}/M_\pi > 1$, an effective field theory approach which neglected the singlet would be somewhat self-defeating; we are interested in the properties of the $0^{++}$ itself, the most novel state in the spectrum.  Therefore, both for practical purposes of doing chiral extrapolations and for the theoretical interest in the combined low energy description of pions plus the singlet scalar, we should seek to compute additional observables which will clarify the form of the low energy EFT in these mass ranges.

\section{Maximal Isospin $\pi \pi$ Scattering \label{sec:three}}

To begin probing the EFT of the eight flavor theory, we study maximal isospin $\pi\pi \rightarrow \pi\pi$ scattering.  In a theory of pions with a light flavor-singlet scalar, one would expect the channel to be dominated by tree level scalar exchange plus the usual four pion vertex, as in the linear sigma model.  This is qualitatively different from $\chi$PT, in which the first correction to the four pion vertex comes from two pion exchange at loop level.  Thus, we expect this channel to provide us nontrivial information about the deviation of the eight flavor theory from QCD-like low energy behavior.  We perform the scattering calculations on some of the same ensembles used in the eight flavor spectrum calculation \cite{8flightsigma2}.  Some of the details of these ensembles are listed in Table~\ref{ensembles}.
\begin{table}
\begin{center}
	
	\begin{tabular}{|| c c c ||} 
		\hline
		Size & $a m_q$ & $(M_\pi / F_\pi)^2$ \\
		\hline
		\hline
		$24^3 \times 48$ & 0.00889 & 19.207(50) \\ 
		\hline
		$32^3 \times 64$ & 0.00750 & 18.303(25)  \\ 
		\hline
		$32^3 \times 64$ & 0.00500 & 18.155(54)  \\ 
		\hline
		$48^3 \times 96$ & 0.00222 & 16.134(43) \\
		\hline
	\end{tabular}
	\caption{Lattice details of ensembles used for $\pi\pi$ scattering calculation.} \label{ensembles}
\end{center}
\end{table}

At finite lattice spacing, an exact $SU(2)$ ``isospin'' symmetry exists, and within each staggered family it may be used to categorize the one particle hadronic states as well as the multi-particle scattering states.  For example, if we denote the staggered fermions by $(\chi_1,\chi_2)$ then we may take $\pi^+ = \chi_2 \epsilon \chi_1$ to source the distance zero Goldstone pion which is the highest weight state in the vector representation of the isospin group.  Here $\epsilon(x) = (-1)^{x+y+z+t}$ denotes the staggered phase corresponding to the $\gamma_5 \times \xi_5$ spin-taste structure \cite{staggered}.  The orientation of the highest weight state within the adjoint multiplet (or $\mathbf{63}$) of the continuum flavor group is a matter of convention, so we can take $\pi^+$ to also be the highest weight state of the larger continuum multiplet.

Therefore to study the maximal isospin scattering channel we consider the s-wave scattering of distance zero (local) Goldstone pions sourced by the operator
\begin{equation}
\pi^+(t) = \sum_{\vec{x}} \bar{\chi_2}(\vec{x},t) \epsilon(\vec{x},t) \chi_1(\vec{x},t) \label{eq:pionop}
\end{equation}
where we have projected onto zero momentum for s-wave scattering.  From this, we construct the simplest two body operator which sources the maximal isospin two pion state.
\begin{equation}
\mathcal{O}_2(t) = \pi^+(t) \pi^+(t+1) \label{eq:i2op}
\end{equation}
We choose to separate the operators by one time slice to avoid projection onto unwanted states due to Fiertz rearrangement identities \cite{i2scattering1}.   We compute the scattering phase shift using Luscher's method \cite{Luscher}.  One computes the total energy of the two pion state from the two point function $\langle \mathcal{O}_2 (t) \mathcal{O}_2^\dagger (0) \rangle$, which is a four point function of one particle interpolating operators.  The energy of interaction of the two pion system is the total energy of the two pion state minus twice the mass of the pion.  The corresponding momentum transfer is given by
\begin{equation}
k^2 = \frac{1}{4}E_{\pi\pi}^2 - M_{\pi}^2
\end{equation}
Once the scattering momentum is known, one may extract the scattering phase shift at momentum transfer $k$ by Luscher's formula
\begin{equation}
k \cot \delta_{l=0}(k)  = \frac{2 \pi}{L} \pi^{-3/2}Z_{00}\left(1,\left (\frac{kL}{2\pi}\right)^2\right)
\end{equation}
where the zeta function is expressed formally as
\begin{equation}
Z_{00}(1;q^2) = \frac{1}{\sqrt{4\pi}} \sum_{n \in \mathbb{Z}^3} \frac{1}{n^2 - q^2}
\end{equation}
From the phase shift, one may make use of the effective range expansion of the scattering phase shift to extract the scattering length and effective range
\begin{equation}
k\cot \delta_0 (k) = \frac{1}{A} + r\frac{M_\pi^2}{2} \left(\frac{k}{M_\pi} \right)^2 + \mathcal{O}\left(\left(\frac{k}{M_\pi} \right)^4\right) 
\end{equation}
where we denote the scattering length by $A$ to distinguish it from the lattice spacing, $a$.

In principle one may construct a large basis of interpolating operators for extracting excited state energies of the two pions scattering.  These excited state energies correspond to scattering states with larger scattering momentum, $k$.  This is particularly important in scattering studies of resonant channels such as the $I=1$ and $I=0$ channels in QCD in which the $\rho$ and the $\sigma$ are resonant, respectively, as one seeks to compute how $\delta_l$ varies with $k$ as $k$ passes through the resonant energy \cite{jlabscattering}.  We are currently working to expand our basis of interpolating operators and extract excited state energies in order to fit to the effective range expansion and compute both $A$ and $r$.  In this work, we focus on only the simplest interpolating operator, Eq.~\ref{eq:i2op}, and we find that the ground state scattering momentum is small enough that the approximation $k\cot \delta_0 (k) \approx \frac{1}{A}$ is justified.


\begin{figure}[t!]
	\begin{center}
		\includegraphics[width=0.5\textwidth]{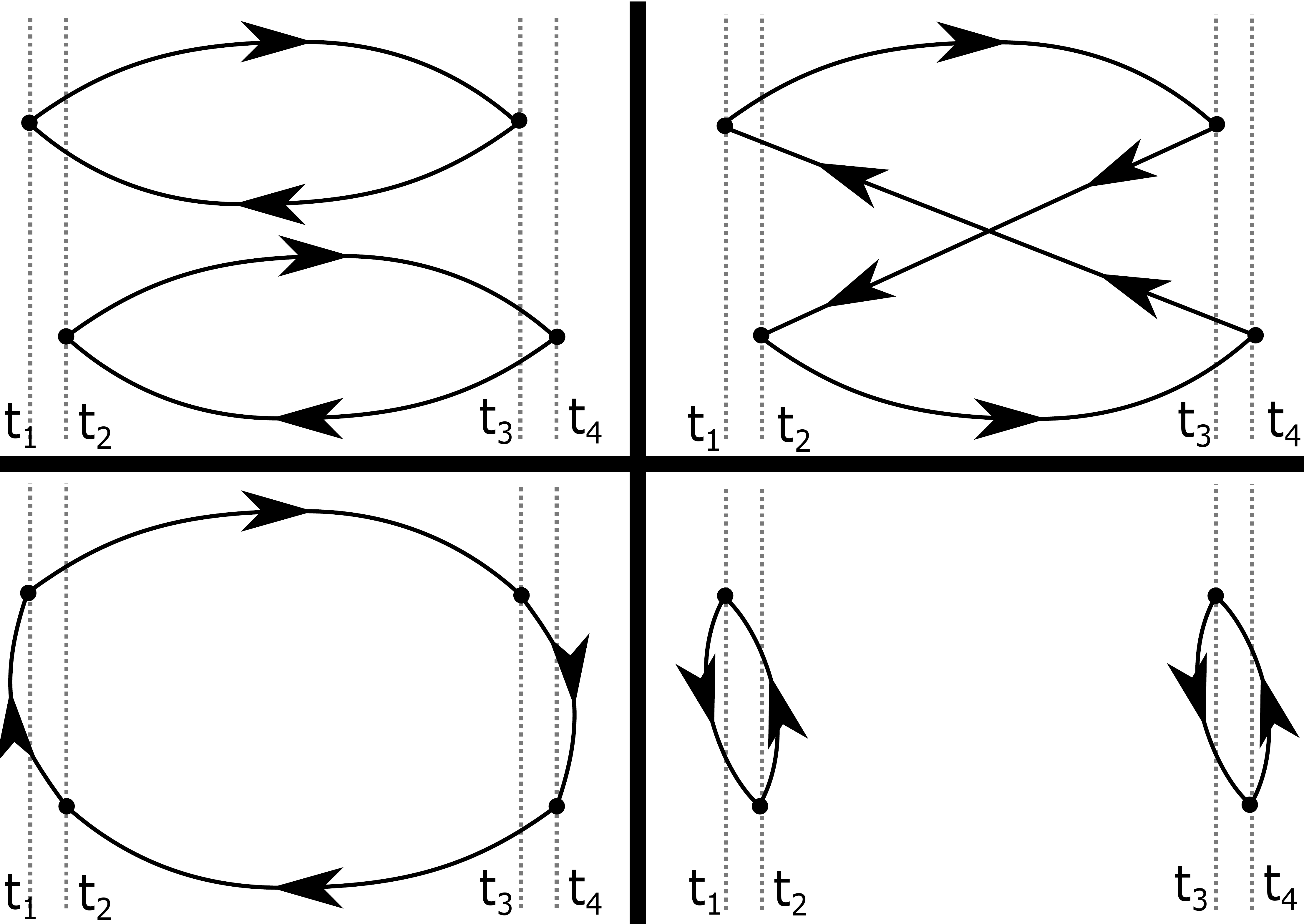}
	\end{center}
	\caption{Wick contractions appearing in $2\rightarrow 2$ scattering processes.  The top left cell is the ``direct'' (D) channel, the top right is the ``crossed'' (C) channel, the bottom left is the ``rectangle'' (R) channel, and the bottom right is the ``vacuum'' (V) channel.  In all diagrams, time flows from left to right.}
	\label{fig:scatt_all}
\end{figure}

The four pion amplitude corresponding to the interpolating operator Eq.~\ref{eq:i2op} is of the form
\begin{eqnarray}
&&\langle {\pi^+}(t_1)^\dagger {\pi^+}(t_2)^\dagger  {\pi^+}(t_3)  {\pi^+}(t_4) \rangle \nonumber \\
&&= \sum_{\vec{x_1}\vec{x_2}\vec{x_3}\vec{x_4}} \epsilon(x_1) \epsilon(x_2) \epsilon(x_3) \epsilon(x_4) \langle \bar{\chi}_1(x_1) \chi_2(x_1) \bar{\chi}_1(x_2) \chi_2(x_2) \bar{\chi}_2(x_3) \chi_1(x_3) \bar{\chi}_2(x_4) \chi_1(x_4) \rangle 
\end{eqnarray}
where we take $t_3,t_4 \gg t_1,t_2$.  The possible Wick contractions are
\begin{align*}
\langle \bar{\chi}_1(x_1) \chi_2(x_1) \bar{\chi}_1(x_2) \chi_2(x_2) \bar{\chi}_2(x_3) \chi_1(x_3) \bar{\chi}_2(x_4) \chi_1(x_4) \rangle \nonumber \\ 
\contraction[2ex]{  = \langle }{ \bar{\chi}_1}{(x_1) \chi_2(x_1) \bar{\chi}_1(x_2) \chi_2(x_2) \bar{\chi}_2(x_3) }{ \chi_1}
\contraction{= \langle \bar{\chi}_1(x_1) }{ \chi_2}{(x_1) \bar{\chi}_1(x_2) \chi_2(x_2)}{ \bar{\chi}_2}
\bcontraction[2ex]{ = \langle \bar{\chi}_1(x_1) \chi_2(x_1)}{ \bar{\chi}_1}{(x_2) \chi_2(x_2) \bar{\chi}_2(x_3) \chi_1(x_3) \bar{\chi}_2(x_4)}{ \chi_1}
\bcontraction{= \langle \bar{\chi}_1(x_1) \chi_2(x_1) \bar{\chi}_1(x_2) }{) \chi_2}{(x_2) \bar{\chi}_2(x_3) \chi_1(x_3)}{ \bar{\chi}_2}
= \langle \bar{\chi}_1(x_1) \chi_2(x_1) \bar{\chi}_1(x_2) \chi_2(x_2) \bar{\chi}_2(x_3) \chi_1(x_3) \bar{\chi}_2(x_4) \chi_1(x_4) \rangle \nonumber \\
\contraction[2ex]{+ \langle}{ \bar{\chi}_1}{(x_1) \chi_2(x_1) \bar{\chi}_1(x_2) \chi_2(x_2) \bar{\chi}_2(x_3) \chi_1(x_3) \bar{\chi}_2(x_4) }{\chi_1}
\contraction{+ \langle \bar{\chi}_1(x_1)}{\chi_2}{(x_1) \bar{\chi}_1(x_2) \chi_2(x_2) \bar{\chi}_2(x_3) \chi_1(x_3) }{ \bar{\chi}_2}
\bcontraction[2ex]{+ \langle \bar{\chi}_1(x_1) \chi_2(x_1)}{\bar{\chi}_1}{(x_2) \chi_2(x_2) \bar{\chi}_2(x_3) }{ \chi_1}
\bcontraction{+ \langle \bar{\chi}_1(x_1) \chi_2(x_1) \bar{\chi}_1(x_2)}{\chi_2}{(x_2) }{\bar{\chi}_2}
+ \langle \bar{\chi}_1(x_1) \chi_2(x_1) \bar{\chi}_1(x_2) \chi_2(x_2) \bar{\chi}_2(x_3) \chi_1(x_3) \bar{\chi}_2(x_4) \chi_1(x_4) \rangle \nonumber \\
\contraction{+ \langle \bar{\chi}_1(x_1) \chi_2(x_1) \bar{\chi}_1(x_2)}{ \chi_2}{(x_2)}{\bar{\chi}_2}
\contraction[2ex]{+ \langle}{\bar{\chi}_1}{(x_1) \chi_2(x_1) \bar{\chi}_1(x_2) \chi_2(x_2) \bar{\chi}_2(x_3)}{ \chi_1}
\bcontraction{+ \langle \bar{\chi}_1(x_1)}{\chi_2}{(x_1) \bar{\chi}_1(x_2) \chi_2(x_2) \bar{\chi}_2(x_3) \chi_1(x_3)}{\bar{\chi}_2}
\bcontraction[2ex]{+ \langle \bar{\chi}_1(x_1) \chi_2(x_1)}{\bar{\chi}_1}{(x_2) \chi_2(x_2) \bar{\chi}_2(x_3) \chi_1(x_3) \bar{\chi}_2(x_4)}{ \chi_1}
+ \langle \bar{\chi}_1(x_1) \chi_2(x_1) \bar{\chi}_1(x_2) \chi_2(x_2) \bar{\chi}_2(x_3) \chi_1(x_3) \bar{\chi}_2(x_4) \chi_1(x_4) \rangle \nonumber  \\
\contraction[2ex]{+ \langle}{\bar{\chi}_1}{(x_1) \chi_2(x_1) \bar{\chi}_1(x_2) \chi_2(x_2) \bar{\chi}_2(x_3) \chi_1(x_3) \bar{\chi}_2(x_4)}{ \chi_1}
\contraction{+ \langle \bar{\chi}_1(x_1) \chi_2(x_1) \bar{\chi}_1(x_2) }{\chi_2}{(x_2) \bar{\chi}_2(x_3) \chi_1(x_3)}{\bar{\chi}_2}
\bcontraction{+ \langle \bar{\chi}_1(x_1) \chi_2(x_1)}{\bar{\chi}_1}{(x_2) \chi_2(x_2) \bar{\chi}_2(x_3)}{ \chi_1}
\bcontraction[2ex]{+ \langle \bar{\chi}_1(x_1)}{ \chi_2}{(x_1) \bar{\chi}_1(x_2) \chi_2(x_2)}{ \bar{\chi}_2}
+ \langle \bar{\chi}_1(x_1) \chi_2(x_1) \bar{\chi}_1(x_2) \chi_2(x_2) \bar{\chi}_2(x_3) \chi_1(x_3) \bar{\chi}_2(x_4) \chi_1(x_4) \rangle \nonumber 
\end{align*}
The first two terms are referred to as the ``direct'' channel, and the second two terms are the ``crossed'' channel.  Employing $\gamma_5$ hermiticity and taking into account the anticommutative properties of the Grassman valued fields, one arrives at
\begin{equation}
\langle {\pi^+}(t_1)^\dagger {\pi^+}(t_2)^\dagger  {\pi^+}(t_3)  {\pi^+}(t_4) \rangle = C_D(13;24) + C_D(14;23) - C_C(1324) - C_C(1423)
\end{equation}
with
\begin{eqnarray}
C_D(i k;j l) = \text{Tr}(G_{x_i x_k}^\dagger G_{x_i x_k} )\text{Tr}_C(G_{x_j x_l}^\dagger G_{x_j x_l} ) \\ 
C_C(i j k l) = \text{Tr}(G_{x_i x_k}G_{x_j x_k}^\dagger G_{x_j x_l} G_{x_i x_l}^\dagger )
\end{eqnarray}
where Tr(...) denotes a color trace as well as a spatial sum over time sheets and $G_{x y}$ is a quark propagator from $x$ to $y$.

Valence quark diagrams in Fig.~\ref{fig:scatt_all} help to visualize the different scattering channels.  The ``rectangle'' and ``vacuum'' diagrams only contribute to pion scattering with nonmaximal isospin, and they tend to be noisier and computationally more expensive \cite{i2scattering1}.  The absence of these diagrams makes maximal isospin scattering a good first channel to investigate.  


\section{Scattering Results and Effective Field Theory Considerations \label{sec:four}}

Now we turn to our results for the scattering analysis outlined in Section~\ref{sec:three} applied to the eight flavor theory.  We appeal to intuition from next-to-leading order $\chi$PT to begin to make sense of our results, and attempt to quantify the applicability of some chiral expansions to our data.  A more thorough effective field theory analysis which incorporates a light $0^{++}$ state is left to a future work. 

Consider the next to leading order expression in chiral perturbation theory with a general $SU(N_F) \times SU(N_F) \rightarrow SU(N_F)$ symmetry breaking pattern \cite{xpt1,xpt2}

\begin{align}
M_\pi^2 &=M^2 \left[ 1 + \frac{ N_F M^2 }{16\pi^2 F^2} \left( 128\pi^2\left( 2L_6^r - L_4^r + \frac{1}{N_F}(2L_8^r - L_5^r)\right) + \frac{1}{N_F^2} \log(M^2/\mu^2)\right)\right]  \label{eq:XPT1}
\\
F_\pi& = F \left[ 1 + \frac{ N_F M^2 }{16\pi^2 F^2} \left( 64\pi^2\left(L_4^r + \frac{1}{N_F}L_5^r\right) -\frac{1}{2} \log(M^2/\mu^2)\right)\right]  \label{eq:XPT2}
\\
M_\pi A &= \frac{-M^2}{16\pi F^2} \left[ 1 + \frac{N_F M^2}{16\pi^2 F^2} \left( -256\pi^2 \left( (1-\frac{2}{N_F})(L_4^r - L_6^r) + \frac{1}{N_F} ( L_0^r + 2L_1^r + 2L_2^r + L_3^r ) \right) \nonumber \right.\right. \\
& \left. \left.- 2\frac{N_F - 1}{N_F^3} + \frac{(2-N_F + 2N_F^2 + N_F^3)}{N_F^3}\log(M^2/\mu^2)  \right) \right] , \label{eq:XPT3}
\end{align}
where $M^2 = 2 B m_q$ as usual.  The loop parameter controlling the convergence of the perturbative expansion, $R \equiv \frac{ N_F M^2 }{16\pi^2 F^2}$, should be approximately less than one if the expansion is to be trusted.  It is enhanced by an explicit $N_F$ factor compared to the QCD value due to the counting of pions in the loop.  In addition, the chiral limit values of $B$ and $F$ are unknown and may change significantly compared to QCD.  

\begin{figure}[t!]
	\begin{center}
		\includegraphics[width=0.6\textwidth]{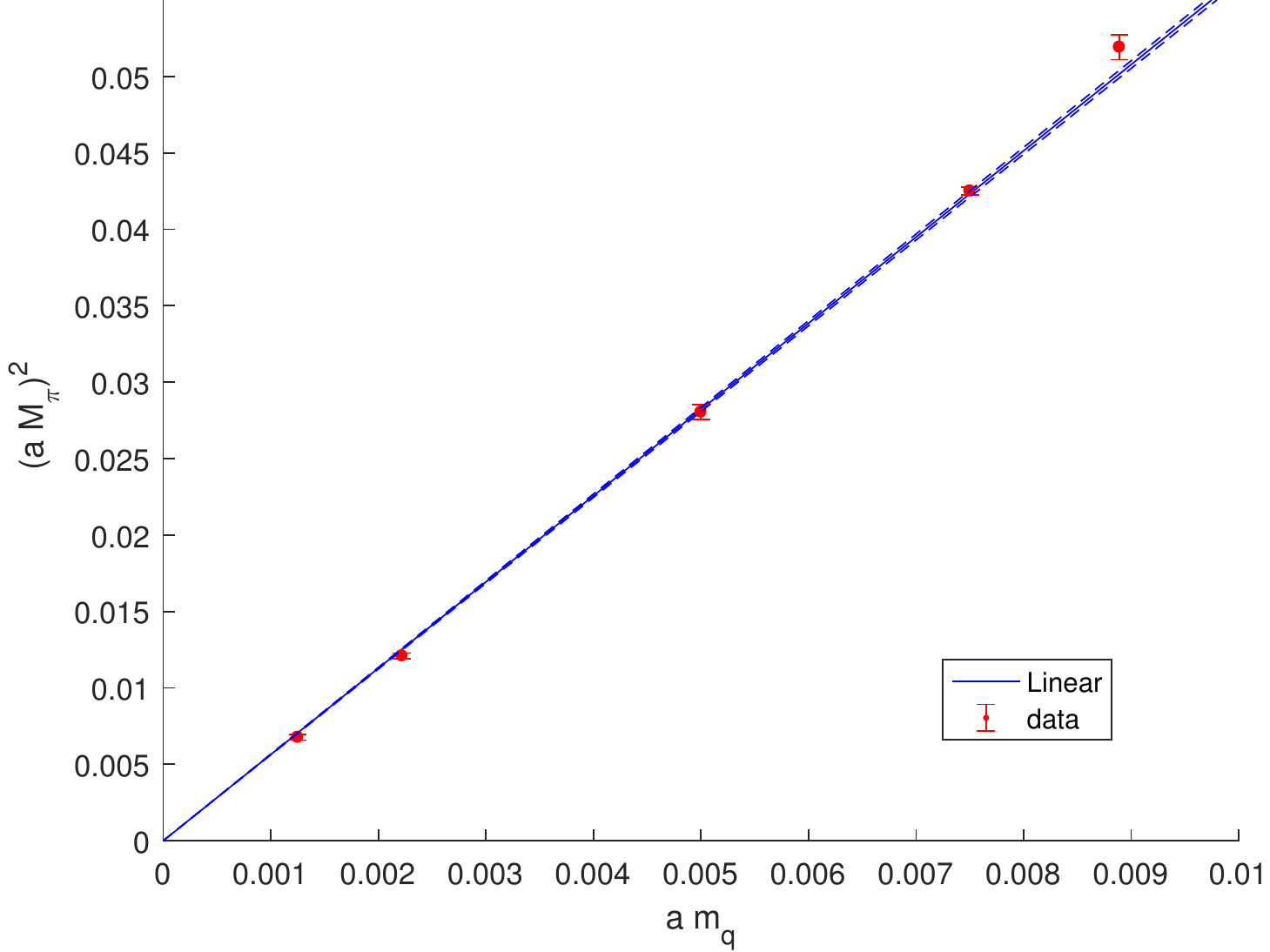}
	\end{center}
	\caption{Leading order $\chi$PT fit to the pion mass.  The solid line is produced by the mean fit parameter values, and the dotted lines denote the $\pm 1\sigma$ uncertainties in the fit parameter.}
	\label{fig:Mpi2_fits}
\end{figure}

The get a sense of the magnitude of the expansion parameter, we attempt to extract $B$ and $F$ from the quark mass dependence of $M_\pi^2$ and $F_\pi$.  In the mass ranges studied, $M_\pi^2$ is highly linear and a LO expression fits well with $\chi^2 / \text{dof} = 3.36$.  We show the fit in Fig.~\ref{fig:Mpi2_fits}.  In lattice units, the low energy constant extracted from the fit is $a B = 5.642(28)$.  The linear behavior of $M_\pi^2$ may be due to the suppression of the log term at NLO by an extra factor of $1/N_F^2$.  In other candidate EFTs, one should see a similar suppression of loop contributions to the pion mass if one hopes to match the observed behavior in the data here.  

\begin{figure}[t!]
	\begin{center}
		\includegraphics[width=0.6\textwidth]{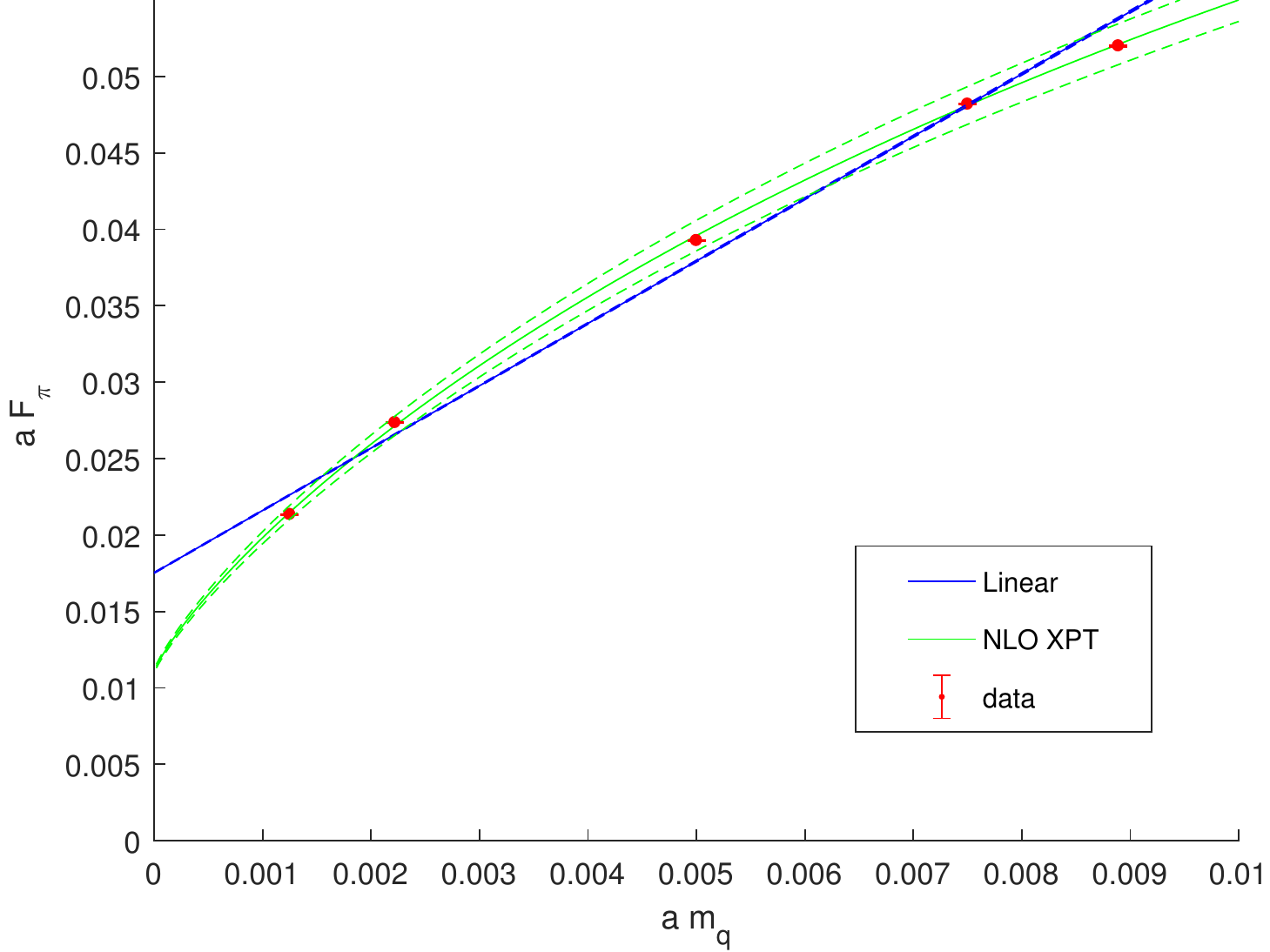}
	\end{center}
	\caption{Linear and NLO $\chi$PT fits to $F_\pi$.  The solid curves are produced by the mean fit parameter values, and the dotted curves denote the $\pm 1\sigma$ uncertainties in the fit parameters. }
	\label{fig:fpi_fits}
\end{figure}

The leading order expression for $F_\pi$, on the other hand, is a constant and clearly does not fit our computed $F_\pi$ values well, which vary significantly through the range of quark masses studied.  Instead, we consider two possible fit functions: a linear extrapolation and an NLO fit.  The results of the fits are shown in Fig.~\ref{fig:fpi_fits}.  For the linear fit, we find the extrapolated value  $a F = 0.017526(31)$ with a chi-square of $\chi^2 / \text{dof} = 1271.33$.  When we fit to the NLO expression we find a smaller extrapolated value of $a F = 0.01109(11)$, with a chi-square of $\chi^2 / \text{dof} = 63.89$.  The large $\chi^2$ values are due to the small error bars on $F_\pi$ which only include statistical errors.  Here, the NLO fit of $F_\pi$ was done independently of the fit to $M_\pi^2$ even though some of the parameters are shared between the two expressions.  We find that a simultaneous NLO fit to $M_\pi$ and $F_\pi$ has a very large $\chi^2$ so we do not record the result here.  Though we note that there is a tendency for the extrapolated $F$ to be slightly larger in a combined fit.  

For the linearly extrapolated value of $F$, we find the expansion parameter $R/(a m_q) = 1861(11)$.  Then for the lightest eight flavor lattices available to us, the expansion parameter is $R(a m_q = 0.00125) = 2.326(14)$ and $R(a m_q = 0.00222) = 4.132(25)$.  For $F$ extrapolated from NLO $\chi$PT, the expansion parameter is $R/(a m_q) = 4648(95)$ yielding  $R(a m_q = 0.00125) = 5.81(12)$ and $R(a m_q = 0.00222) = 10.32(21)$ on the lightest eight flavor ensembles.  So, for $\chi$PT to be valid we will likely need to go to lighter bare quark masses by a factor of at least 2 to 5.  If our attempts at chiral extrapolations have overestimated the chiral value of $F$, then the required quark masses would be lighter still.  The applicability of chiral perturbation theory at existing quark masses does not look promising, but this is not to say that calculations in these mass ranges are without value.  On the contrary, we are interested in dynamics which incorporate both the pions and the $0^{++}$.  So, fits of low energy expressions from other candidate EFTs---which are expected to have a larger radius of convergence due to the incorporation of the scalar---will likely reveal more.  Recall that when the chiral Lagrangian is derived from a linear sigma model, the radius of convergence is set by the sigma mass.  At light enough quark masses when $M_{0^{++}} / M_\pi > 1$ there should exist a range of energies $q^2,M_\pi^2 < M_{0^{++}}^2$ in which chiral perturbation theory is valid.  Conversely, the quark mass at which our estimate of the chiral expansion parameter $R$ becomes less than one could be used as an estimate for where we expect $M_{0^{++}}$ and $M_\pi$ to split, a phenomenon which has not yet been observed in the eight flavor theory.

\begin{figure}[t!]
	\begin{center}
		\includegraphics[width=0.6\textwidth]{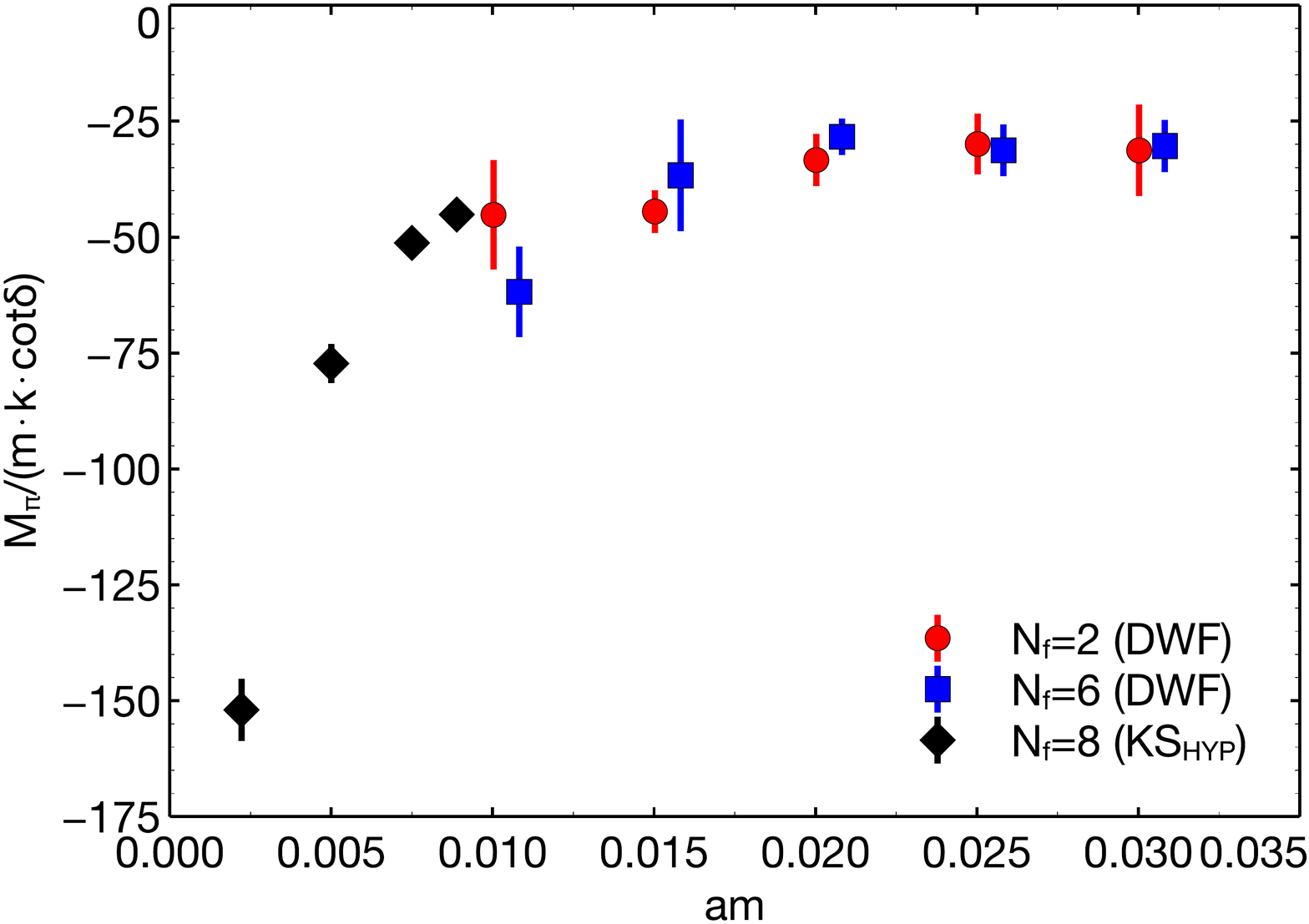}
	\end{center}
	\caption{Scattering phase shift plotted against bare quark masses for $N_F=8$ Kogut-Susskind nHYP-smeared fermions ($\text{KS}_\text{HYP}$).  Data from \cite{WW} is included of $N_F = 2$ and $N_F = 6$ domain wall fermions (DWF) for comparison.  Note that because dimensionful quantities are given in lattice units which vary between $N_F=8$, $N_F = 6$, and $N_F=2$, the comparison between them is only qualitative.}
	\label{fig:scatteringlength2}
\end{figure}

Bearing in mind our assessment of the viability of $\chi$PT in these mass ranges, let us now turn to our results for the maximal isospin scattering length.  Fig.~\ref{fig:scatteringlength2} shows $M_\pi / (m_q k\cot(\delta)) \approx M_\pi A / m_q$ which corresponds to the $\chi$PT expansion Eq.~\ref{eq:XPT3} with the leading order dependence on $m_q$ divided out.  We include data from a previous study of $N_F=2$ and $N_F=6$ for comparison \cite{WW}.  At sufficiently small quark masses, $\chi$PT should apply and the quantity should approach a constant.  The $N_F = 2$ points are statistically consistent with a constant, while the $N_F=6$ points begin to show some curvature and our $N_F=8$ points show very striking curvature.  As $N_F$ increases, the range of $a m_q$ for which we expect $\chi$PT to converge and the plot to behave like a constant decreases rapidly.  Though, we do expect that the $N_F=8$ data will flatten out at some sufficiently small value of $a m_q$.  If we use our extrapolated values for $a B$ and $a F$ to estimate the chiral limit value, we find $M_\pi A / (a m_q) = -730.8 (4.5)$ for the linearly extrapolated $F$ value and $M_\pi A / (a m_q) = -1825(37)$ for the $F$ value extracted from NLO $\chi$PT.  One sees that the $N_F=8$ data in Fig.~\ref{fig:scatteringlength2} is still far from these values even at the lightest quark masses.


\begin{figure}[t!]
	\begin{center}
		\includegraphics[width=0.6\textwidth]{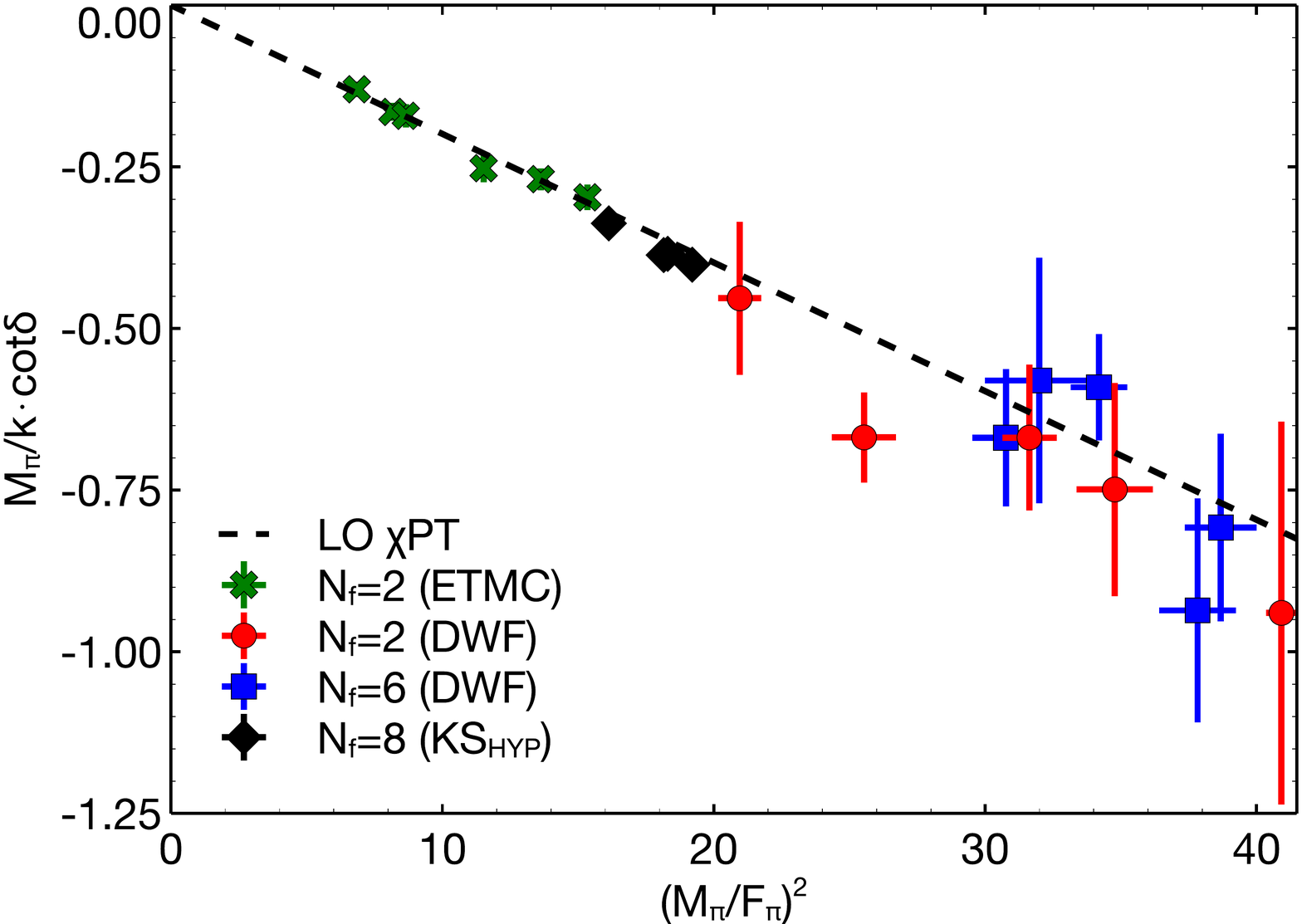}
	\end{center}
	\caption{Scattering phase shift plotted against measured values of $(M_\pi / F_\pi)^2$.  $N_F = 2$ and $N_F = 6$ domain wall fermions data from \cite{WW} (DWF) and $N_F = 2$ data at lighter quark masses from \cite{nf2scattering} (ETMC) are included for comparison. }
	\label{fig:scatteringlength1}
\end{figure}

We also plot the scattering length against the measured values of $(M_\pi / F_\pi)^2$ in Fig.~\ref{fig:scatteringlength1}.  To compare to $\chi$PT, one must reexpand the expression for the scattering length in terms of the physical values of $M_\pi$ and $F_\pi$.  
\begin{align}
M_\pi A & = - \frac{M_\pi^2}{16\pi F_\pi^2} \left( 1 + \frac{M_\pi^2}{16 \pi^2 F_\pi^2} \left( -256 \pi^2 \left( L_0^r + 2 L_1^r + 2 L_2^r + L_3^r - 2 L_4^r -L_5^r + 2 L_6^r  + L_8^r \right) \right.\right. \nonumber \\
& \left. \left.  - \frac{ 2(N_F - 1) }{N_F^2} + \frac{2 (1 - N_F + N_F^2)}{N_F^2} \log\left( \frac{M_\pi^2}{\mu^2} \right) \right)  \right)
\end{align}
In this expansion for $M_\pi A$, it appears that the expansion parameter that controls the convergence is diminished by a factor of $N_F$.  However, if one looks at a wider range of observables in this expansion scheme the factor of $N_F$ is not missing elsewhere.  We see a strong agreement with leading order behavior in Fig.~\ref{fig:scatteringlength1} when plotted against physical quantities, but we do not take this as an indication that chiral perturbation theory is converging.  Rather we consider this a peculiarity of this particular expansion for this particular observable, $M_\pi A$, just as $M_\pi^2$ is anomalously well described by the leading order $\chi$PT term when expanded in bare quantities.

\section{Conclusions \label{sec:five}}

With this work, we have begun a new line of inquiry about the low energy dynamics of walking gauge theory.  While previous works have sought mainly to establish the walking behavior and to calculate the masses of the low lying states, we have started to investigate how states interact in this regime---that is, to establish the appropriate low energy effective theory of the light states.  In QCD, one typically considers the $\sigma$ mass as setting the radius of convergence of chiral perturbation theory.  The lightness of the flavor singlet scalar in the eight flavor theory, which is now expected to be a generic feature of walking models, suggests some new EFT should apply which encapsulates the dynamics of the Goldstone bosons and the scalar (or possibly, scalars) up to the next heaviest state, the $\rho$ mass.

Our assessment of $M_\pi^2$, $F_\pi$, and $M_\pi A$ seems to suggest that chiral perturbation theory does not describe the low energy physics well in the mass ranges investigated.  However, it has given us some intuition about the distance from the chiral limit at which we might expect to see a splitting of the approximate degeneracy of $M_{0^{++}}$ and $M_\pi$.  In upcoming work, we will consider a wider effective field theory analysis in which we test candidate theories which do incorporate a light scalar state such as the linear sigma model and a variety of dilatonic effective field theories \cite{Matsuzaki:2013eva,EFT1,EFT2,EFT3,EFT4}.

\section*{Acknowledgments}This research used resources of the Argonne Leadership Computing Facility, which is a DOE Office of Science User Facility supported under Contract DE-AC02-06CH11357.  The authors thank the Lawrence Livermore National Laboratory (LLNL) Multiprogrammatic and Institutional Computing program for Grand Challenge allocations and time on the LLNL BlueGene/Q supercomputer, along with funding from LDRD 13-ERD-023.  Additional numerical analyses were carried out on clusters at LLNL, Boston University and Fermilab.  G.T.F. was supported by NSF grant PHY-1417402.  The authors would like to specifically acknowledge collaborators Evan Weinberg, Xiao-Yong Jin, and Enrico Rinaldi for contributions to code development, running computations at ANL, analysis, and help generating figures for this paper.  A.D.G. thanks James Ingoldby and Thomas Appelquist for many useful discussions regarding effective field theory.


\begin{thebibliography}{99}
	
	\bibitem{conformalwindow} 
	T.~Appelquist, G.~T.~Fleming and E.~T.~Neil,
	Phys.\ Rev.\ Lett.\  {\bf 100}, 171607 (2008)
	Erratum: [Phys.\ Rev.\ Lett.\  {\bf 102}, 149902 (2009)]
	doi:10.1103/PhysRevLett.100.171607
	[arXiv:0712.0609 [hep-ph]].
	
	\bibitem{conformalwindow2}
	T.~Appelquist {\it et al.} [LSD Collaboration],
	Phys.\ Rev.\ Lett.\  {\bf 104}, 071601 (2010)
	doi:10.1103/PhysRevLett.104.071601
	[arXiv:0910.2224 [hep-ph]].
	
	\bibitem{Pica:2017gcb} 
	Pica, Claudio
	PoS LATTICE {\bf 2016}, 015 (2016)
	[arXiv:1701.07782 [hep-lat]].
	
	\bibitem{Sparameter}
	T.~Appelquist {\it et al.} [LSD Collaboration],
	Phys.\ Rev.\ Lett.\  {\bf 106}, 231601 (2011)
	doi:10.1103/PhysRevLett.106.231601
	[arXiv:1009.5967 [hep-ph]].
	
	\bibitem{8flightsigma1}
	Y.~Aoki {\it et al.} [LatKMI Collaboration],
	Phys.\ Rev.\ D {\bf 89}, 111502 (2014)
	doi:10.1103/PhysRevD.89.111502
	[arXiv:1403.5000 [hep-lat]].
	
	\bibitem{8flightsigma2}
	T.~Appelquist {\it et al.},
	Phys.\ Rev.\ D {\bf 93}, no. 11, 114514 (2016)
	doi:10.1103/PhysRevD.93.114514
	[arXiv:1601.04027 [hep-lat]].
	
	\bibitem{8flightsigma3}
	Y.~Aoki {\it et al.} [LatKMI Collaboration],
	arXiv:1610.07011 [hep-lat].
	
	\bibitem{750GeV1}
	Tech. Rep. ATLAS-CONF-2015-081, CERN, Geneva (2015), http://cds.cern.ch/record/2114853
	\bibitem{750GeV2}
	Tech. Rep. CMS-PAS-EXO-15-004, CERN, Geneva (2015),  https://cds.cern.ch/record/2114808
	
	\bibitem{750model1}
	K.~Harigaya and Y.~Nomura,
	Phys.\ Lett.\ B {\bf 754}, 151 (2016)
	doi:10.1016/j.physletb.2016.01.026
	[arXiv:1512.04850 [hep-ph]].
	
	\bibitem{8frunning}
	A.~Hasenfratz, D.~Schaich and A.~Veernala,
	JHEP {\bf 1506}, 143 (2015)
	doi:10.1007/JHEP06(2015)143
	[arXiv:1410.5886 [hep-lat]].
	
	\bibitem{Fodor:2015baa} 
	Fodor, Zoltan, K.~Holland, J.~Kuti, S.~Mondal, D.~Nogradi and C.~H.~Wong,
	JHEP {\bf 1506}, 019 (2015)
	doi:10.1007/JHEP06(2015)019
	[arXiv:1503.01132 [hep-lat]].
	
	\bibitem{8fDW}
	T.~Appelquist {\it et al.} [LSD Collaboration],
	Phys.\ Rev.\ D {\bf 90}, no. 11, 114502 (2014)
	doi:10.1103/PhysRevD.90.114502
	[arXiv:1405.4752 [hep-lat]].
	
	\bibitem{addition1}
	R.~C.~Brower, A.~Hasenfratz, C.~Rebbi, E.~Weinberg and O.~Witzel,
	Phys.\ Rev.\ D {\bf 93}, no. 7, 075028 (2016)
	doi:10.1103/PhysRevD.93.075028
	[arXiv:1512.02576 [hep-ph]].
	
	\bibitem{addition2}
	Hasenfratz, Anna, Claudio Rebbi and Oliver Witzel,
	arXiv:1609.01401 [hep-ph].

	\bibitem{Yamawaki:1985zg} 
	K.~Yamawaki, M.~Bando and K.~i.~Matumoto,
	Phys.\ Rev.\ Lett.\  {\bf 56}, 1335 (1986).
	doi:10.1103/PhysRevLett.56.1335
	
	\bibitem{Bando:1986bg} 
	M.~Bando, K.~i.~Matumoto and K.~Yamawaki,
	Phys.\ Lett.\ B {\bf 178}, 308 (1986).
	doi:10.1016/0370-2693(86)91516-9

	\bibitem{TAdilaton}
	T.~Appelquist and Y.~Bai,
	Phys.\ Rev.\ D {\bf 82}, 071701 (2010)
	doi:10.1103/PhysRevD.82.071701
	[arXiv:1006.4375 [hep-ph]].
	
	\bibitem{2TeV1}
	Aad, Georges, et.al. [ATLAS Collaboration], arXiv:1506.00962 [hep-ex]
	\bibitem{2TeV2}
	CMS Collaboration [CMS Collaboration], CMS-PAS-EXO-14-010.
	
	\bibitem{staggered}
	D.~Daniel and S.~N.~Sheard,
	Nucl.\ Phys.\ B {\bf 302}, 471 (1988).
	doi:10.1016/0550-3213(88)90211-8
	
	\bibitem{i2scattering1}
	Z.~Fu,
	Phys.\ Rev.\ D {\bf 87}, no. 7, 074501 (2013)
	doi:10.1103/PhysRevD.87.074501
	[arXiv:1303.0517 [hep-lat]].
	
	\bibitem{Luscher}
	M.~Luscher,
	Nucl.\ Phys.\ B {\bf 354}, 531 (1991).
	doi:10.1016/0550-3213(91)90366-6
	
	\bibitem{jlabscattering}
	R.~A.~Briceno, J.~J.~Dudek, R.~G.~Edwards and D.~J.~Wilson,
	Phys.\ Rev.\ Lett.\  {\bf 118}, no. 2, 022002 (2017)
	doi:10.1103/PhysRevLett.118.022002
	[arXiv:1607.05900 [hep-ph]].
	
	\bibitem{xpt1}
	J.~Bijnens and J.~Lu,
	JHEP {\bf 0911}, 116 (2009)
	doi:10.1088/1126-6708/2009/11/116
	[arXiv:0910.5424 [hep-ph]].
	
	\bibitem{xpt2}
	J.~Bijnens and J.~Lu,
	JHEP {\bf 1103}, 028 (2011)
	doi:10.1007/JHEP03(2011)028
	[arXiv:1102.0172 [hep-ph]].
	
	\bibitem{WW}
	T.~Appelquist {\it et al.},
	Phys.\ Rev.\ D {\bf 85}, 074505 (2012)
	doi:10.1103/PhysRevD.85.074505
	[arXiv:1201.3977 [hep-lat]].
	
	\bibitem{nf2scattering}
	X.~Feng, K.~Jansen and D.~B.~Renner,
	Phys.\ Lett.\ B {\bf 684}, 268 (2010)
	doi:10.1016/j.physletb.2010.01.018
	[arXiv:0909.3255 [hep-lat]].
	
	\bibitem{Matsuzaki:2013eva} 
	S.~Matsuzaki and K.~Yamawaki,
	Phys.\ Rev.\ Lett.\  {\bf 113}, no. 8, 082002 (2014)
	doi:10.1103/PhysRevLett.113.082002
	[arXiv:1311.3784 [hep-lat]].
	
	\bibitem{EFT1}
	M.~Golterman and Y.~Shamir,
	Phys.\ Rev.\ D {\bf 94}, no. 5, 054502 (2016)
	doi:10.1103/PhysRevD.94.054502
	[arXiv:1603.04575 [hep-ph]].
	
	\bibitem{EFT2}
	J.~Soto, P.~Talavera and J.~Tarrus,
	Nucl.\ Phys.\ B {\bf 866}, 270 (2013)
	doi:10.1016/j.nuclphysb.2012.09.005
	[arXiv:1110.6156 [hep-ph]].
	
	\bibitem{EFT3}
	M.~Hansen, K.~Langaeble and F.~Sannino,
	arXiv:1610.02904 [hep-ph].
	
	\bibitem{EFT4}
	Appelquist, Thomas, James Ingoldby, and Maurizio Piai.  "Pion-Dilaton EFT and Lattice Data."  (in preparation).
	
\end{thebibliography}
\end{document}